\documentclass[letterpaper, 10 pt, journal, twoside]{ieeetran}

\IEEEoverridecommandlockouts                              
\usepackage{amssymb}                    \usepackage{bm}                                   
\usepackage{algorithm2e}
\usepackage{ifpdf}
\usepackage{amsmath}
\usepackage{amsfonts}
\usepackage{mathrsfs}
\usepackage{color}
\usepackage{graphicx}
\usepackage{subfigure}
\usepackage{lineno}
\usepackage{cite}
\usepackage{enumerate}
\usepackage{epsfig}
\DeclareMathOperator{\argmin}{argmin}

\newcommand{\cov}{{\rm{cov}}}

\newcommand{\Eb}{\mathbf{E}}

\newcommand{\RBB}{\mathbb{R}}

\newcommand{\AC}{\mathcal{A}}

\newcommand{\DC}{\mathcal{D}}
\newcommand{\EC}{\mathcal{E}}
\newcommand{\FC}{\mathcal{F}}

\newcommand{\IC}{\mathcal{I}}

\newcommand{\PC}{\mathcal{P}}
\newcommand{\QC}{\mathcal{Q}}

\newcommand{\SC}{\mathcal{S}}

\newcommand{\UC}{\mathcal{U}}

\newcommand{\WC}{\mathcal{W}}
\newcommand{\NC}{\mathcal{N}}
\newcommand{\XC}{\mathcal{X}}

\newtheorem{theorem}{Theorem}
\newtheorem{definition}{Definition}

\newtheorem{lemma}{Lemma}
\newtheorem{problem}{Problem}     
\newtheorem{remark}{Remark}
\newtheorem{proposition}{Proposition}
\newtheorem{assumption}{Assumption}

\newcommand{\eq}{&\hspace{-0.5em}=\hspace{-0.5em}&}
\newcommand{\tr}{{\rm{tr}}}

\title{
Optimal Transmission Power Scheduling for Networked Control System under DoS Attack}

\author{ Siyi Wang, Yulong Gao and Sandra Hirche 
\thanks{*This work was funded
by the Federal Ministry of Education
and Research of Germany in the programme of “Souver{\"an}. Digital. Vernetzt. ” under the joint project 6G-life (Project ID: 16KISK002).}
\thanks{Siyi Wang and Sandra Hirche are with the Chair of Information-oriented Control (ITR), Technical University of Munich, 80333, Munich, Germany. Email:         {\tt\small         \{siyi.wang,hirche\}@tum.de}       }
\thanks{Yulong Gao is with the Department of Electrical and Electronic Engineering, Imperial College London, SW7 2AZ, London, UK.  Email: {\tt\small yulong.gao@imperial.ac.uk}} 
}

\graphicspath{{figures/}}

\begin{document}

\maketitle
\thispagestyle{empty}
\pagestyle{empty}

\begin{abstract}
Designing networked control systems that are reliable and resilient against adversarial threats, is essential for ensuring the security of cyber-physical systems.  
This paper addresses the communication-control co-design problem for networked control systems under denial-of-service (DoS) attacks. 
In wireless channels, a transmission power scheduler periodically determines the power level for sensory data transmission. Yet DoS attacks render data packets unavailable by disrupting the communication channel.
This paper co-designs the control and power scheduling laws in the presence of DoS attacks and aims to minimize the sum of regulation control performance and transmission power consumption. Both finite- and infinite-horizon discounted cost criteria are addressed. By delving into the information structure between the controller and the power scheduler under attack, the original co-design problem is divided into two subproblems that can be solved individually without compromising optimality. The optimal control is shown to be certainty equivalent, and the optimal transmission power scheduling is solved using a dynamic programming approach. Moreover, in the infinite-horizon scenario, we analyze the performance of the designed scheduling policy and develop an upper bound of the total costs. Finally, a numerical example is provided to demonstrate the theoretical results.  
\end{abstract}

\section{Introduction}\label{sec:introduction}

Cyber-physical systems are systems that integrate sensors, controllers, and actuators to collaborate over a communication network for regulating and optimizing the behavior of a dynamic system \cite{zhang2001stability,jiang2023monitoring}. Its applications include robotics \cite{yan2013survey}, smart grids \cite{singh2014stability}, and intelligent vehicle \cite{li2017platoon}. 
Networked systems generally assume that sensory data is measured and transmitted periodically to update control signals \cite{molin2014suboptimal}. 
However, transmitting data over a communication network is generally costly due to the limited battery energy and communication channel bandwidth. This fact motivates us to co-design the control and communication strategies ensuring that the valuable sensory data is efficiently transmitted to the controller, thereby improving
overall system performance. 

Due to increased connectivity,  cyber-physical systems are suffering from cyber threats. 
Two most common cyber attacks are deception attacks and denial-of-service (DoS) attacks \cite{amin2009safe}. Deception attacks degenerate the system performance by maliciously modifying the information contained in transmitted data \cite{teixeira2012attack,zhu2023secure,xu2024output}. 
DoS attacks render data packets unavailable by jamming communication channels \cite{dai2023robust, feng2020networked,zhang2024optimal}. It is more destructive compared to deception attacks when facing threats from large-scale and persistent attacks.  
Considering the detrimental impact of cyberattacks on control performance, proactive risk management strategies are essential to ensure the resilience of networked control systems.

Transmission power scheduling means that the transmission device dynamically adjusts power levels according to the changing network conditions to fulfill requirements of cyber-physical systems \cite{gatsis2014optimal,ren2017infinite}. Compared to the traditional event-based triggering, power scheduling allows for continuous optimization of energy usage,  which is crucial for battery-powered devices and energy-constrained environments. Increasing transmission power levels typically enhances communication link reliability.  As a consequence, the power scheduler envisions a tradeoff between energy usage and improved control performance. 
In networked control systems, power scheduling is typically addressed by selecting a power level from a predetermined finite set to achieve an optimization objective, such as  \cite{li2019multi,xu2023optimal}. Another common approach involves pre-defining a specific scheduling policy and searching for the optimal scheduling parameters by solving associated optimization problems, such as \cite{wu2015data,wu2020mean}. However, these methods often constrain the scheduling policy's structure, leading to suboptimal solutions.  In contrast, \cite{gatsis2014optimal,ren2017infinite,soleymani2021feedback} do not fix the structure of scheduling policies. Instead, they focus on finding the optimal mapping from the system state and communication channel conditions to the power scheduling decision. More specifically, \cite{ren2017infinite} investigates the optimal power scheduling strategy for remote estimation over a fading channel. Similarly, \cite{gatsis2014optimal,soleymani2021feedback}  investigate the joint co-design of power scheduling and control, aiming to minimize the long-term transmission energy and control cost.  However, the above works do not consider the impact of cyber attacks. 

When addressing attacks in networked control systems, this work differs from existing works that focus on the energy allocation of the attacker \cite{ding2018attacks,zhang2018denial,huang2022learning} or joint energy allocation design of the attacker and the power scheduler \cite{li2016sinr}. Instead, we focus on the optimal power scheduler design under DoS attack.  The relating works are \cite{li2015jamming,ding2017multi}, where  \cite{li2015jamming}  
investigate the power allocation under the DoS attack to minimize the long-term mean square error covariance and propose a variance-based scheduling mechanism. However, variance-based scheduling does not utilize real-time innovations when making decisions and thus is generally outperformed by state-based scheduling \cite{wu2013can}. Moreover, \cite{ding2017multi} investigates a multi-channel schedule for remote estimation under DoS attack, where the power is chosen from a pre-determined level set. In contrast, our scheduling method chooses the power values from a continuous real-valued domain, which allows for greater flexibility in design.   
Additionally, in real-world applications, current rewards are typically valued more than future rewards. Thus, this article aims to obtain a co-design strategy that minimizes the expectation of a discounted cumulated cost.

In this work, we propose a framework to jointly co-design the control law and power scheduler for the networked control system under DoS attack.
We consider a signal-to-interference-plus-noise ratio (SINR)-based network model \cite{li2016sinr}, where the transmission success probability is affected by the attack energy and the transmission power level chosen by the scheduler. 
The contribution of this work is summarized as follows. 
For both the finite- and infinite-horizon cases, the optimal co-design of control and scheduling is shown to be separable, given that the knowledge about the attack energy is symmetric between the controller and the scheduler. The optimal control law is shown to be certainty equivalent. We apply the dynamic programming approach to the remaining sequential decision problem. In the finite-horizon case, we provide the analytical solution of the optimal state-based power scheduling design and its greedy version that simplifies computation. In the infinite-horizon case, we solve the corresponding Bellman equation on bounded Borel state space with discounted cost and derive the optimal stationary power scheduling policy. Further, we establish an upper bound on the total cost achieved by the designed power scheduler.

The remainder of this article is structured as follows: Section~\ref{sec:preliminaries} introduces preliminaries. Section~\ref{sec:finite} and Section~\ref{sec:infinite} present the main result on optimal co-design of control and transmission power scheduling in finite- and infinite-horizon cases, respectively. Section~\ref{sec:simulation} presents numerical simulations. Section~\ref{sec:conclusion} concludes this work. 

\textbf{Notations}. 
Denote by $\mathbb{R}$ and $\mathbb{R}^n$ the set of real numbers and the set of the $n$-tuples of real numbers, respectively.  Denote by $\mathbb{N}$ the set of nonnegative integers. 
For $x, y \in \mathbb{N}$ and $x \le y$, the set $\mathbb{N}_{[x,y]}$ denotes $\{z \in \mathbb{N}|x \le z \le y\}$.
Denote $x_{0:k}$ as the history of state $x_{t}$ during time $t \in \mathbb{N}_{[0,k]}$. 
For a random variable $X$,  $X \sim \DC$ implies that $X$ is distributed according to the distribution $\DC$.  
Denote $\Eb[\cdot]$, $\Eb[\cdot\mid\cdot]$ and $\cov[\cdot]$ as the expectation, the conditional expectation and the covariance of the random variable, respectively. 
   If the random variable $x$ follows a normal distribution with the mean of $a$ and the covariance of $\Sigma$, we write $ x \sim \NC(a, \Sigma)$.
Denote by $\mu_{\Sigma}(X) =  \frac{1}{(2 \pi)^{\frac{n}{2}} \vert \Sigma  \vert^{\frac{1}{2}}} \exp (-\frac{(X-a)^{\top}\Sigma^{-1}(X-a)}{2})$  the probability density function (p.d.f) of a $n$-dimension random vector $X \sim \NC(a,\Sigma)$. 
Denote the spectral radius of $A$ as $\rho(A)$.

\section{Preliminaries}\label{sec:preliminaries}

\begin{figure}[t]
    \centering
    \includegraphics[width=0.4\textwidth]{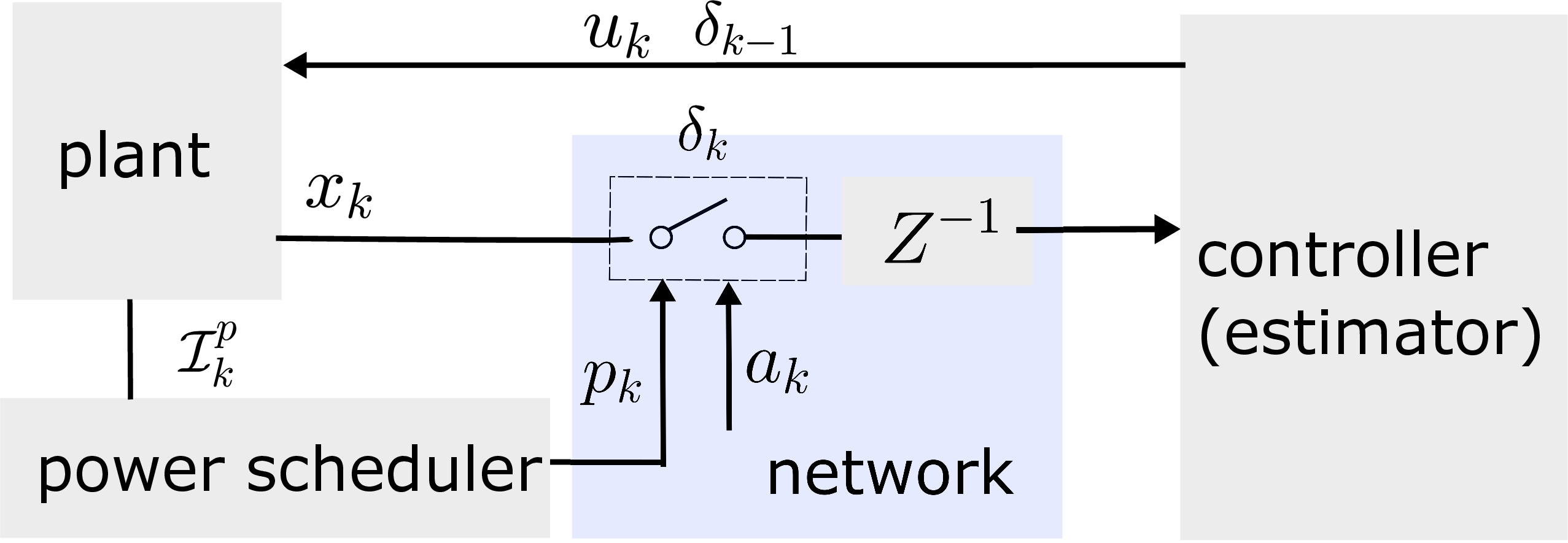}
    \centering 
    \caption{Networked control system with transmission power scheduler.}
    \label{fig:system architecture}
\end{figure}

We consider the energy-constrained feedback control system over the communication network, as illustrated in Fig.~\ref{fig:system architecture}. The scheduler periodically determines the power to transmit sensory data, which affects the transmission success probability of the packets. The controller generates the control signal based on the remote estimate.  
\subsection{System model}
The discrete-time stochastic dynamical system to be controlled is described as
\begin{eqnarray}\label{eq:plant}
x_{k+1}  \eq A x_{k}  +B u_{k}  +w_{k},  
\end{eqnarray}
where  $x_k \in \mathbb{R}^{n}$ and $u_{k}\in \mathbb{R}^{m}$ are the state vector and the control force, respectively. The system matrices are given by $A\in \mathbb{R}^{n\times n}$, $B \in \mathbb{R}^{n\times m}$, where the pair $(A,B)$ is controllable. The process noise $w_k \in \RBB^{n} \sim \NC(0, W)$ is assumed to be independent identically distributed (i.i.d.)  Gaussian processes with zero mean and positive semi-definite variance $W$. The initial state  $x_0 \sim \NC(\bar{x}_{0}, X_{0}) $  is a random vector that is statistically independent of $w_k$ for all $k$.

\subsection{Network model}
The sensor located at the plant side periodically accesses the system states, as in Fig.~\ref{fig:system architecture}. The power-constrained scheduler determines the power $p_k = \pi(\IC_k^p) \in \EC$ used to send out packets at time $k$, where $\pi$ and $\IC_k^p$ denote the power scheduling law and the locally available information set of the power scheduler, respectively. Moreover, denote $\EC = [0,p_{\max}]$ as the admissible transmission power set with  $p_{\max}$ being the maximum transmission power.

Consider an additive white Gaussian noise (AWGN) channel using quadrature amplitude modulation \cite{li2016sinr}. In the presence of a DoS interference attacker,  the communication channel is modeled as ${\rm SINR} = p_k/ 
(a_k + \sigma^2) $, where
$\sigma^2$  is the additive white Gaussian noise power, and $a_k$ is the interference power from the attacker \cite{proakis2008digital}. The following assumptions are for the attack energy. 
\begin{assumption}\label{ass:attack property}
\begin{enumerate}
\item The attack energy $a_k \in \mathbb{R}$, for $k \in \mathbb{N}$, is a i.i.d. random process with a distribution $\DC_a$.     
\item The attack energy  $a_k \in \SC$ with $\SC:=[0,a_{\max}]$ and $a_{\max}$ being a nonnegative scalar. 
\item The random variable $a_k$, $k \in\mathbb{N}$, is independent of process noise $w_k$, for all $k$, and the initial state $x_0$. 
\end{enumerate}
\end{assumption}
The attack energy $a_k$ can be interpreted as a channel fading parameter that encompasses unpredictable variations in the wireless channel \cite{gatsis2014optimal}. It can be measured using real-time monitoring systems. For instance, telecom networks measure DoS attack energy by evaluating traffic volumes, bandwidth consumption, and processing loads on network elements.
The subsequent sections will discuss control and power scheduling co-design, considering known and unknown attack energy scenarios, respectively.

Consider a random binary process $\delta_{k} \in \{ 0,1\}$, where $\delta_k = 1$ denotes the packet is transmitted successfully, and $\delta_k=0$ otherwise.  
The packet dropout probability is affected by the transmission power and attack energy:  
\begin{eqnarray}\label{eq:drop prob}
q_k=
\Pr(\delta_{k}=0 \vert p_{k},a_k) = 2 Q_f\bigg( \sqrt{ \frac{\alpha p_k}{a_k + \sigma^2}}\bigg),  
\end{eqnarray}
where $\alpha$ is a communication channel parameter and $Q_f(\cdot)$ is the tail function of the standard normal distribution \cite{li2016sinr}.  Moreover, $q_k \in \QC:=[2Q_f\big( \sqrt{  \alpha p_{\max}/\sigma^2}\big),1]$.
Note that a higher transmission power indicates a lower packet dropout probability and vice versa.  
Assume that the network induces a one-step delay. 
At time $k$, the packet arrives at the remote side is  $z_k = x_{k-1}$ if $\delta_{k-1} = 1$, and $z_k = \emptyset$ otherwise, 
with $z_0 = \emptyset$. This assumption is widely used to facilitate the sequential decision processes between the
scheduler and the controller, see \cite{soleymani2022value,soleymani2021feedback}. More detailed discussion will be provided in Lemma~\ref{lemma:CE}.  
The remote estimator is given as
\begin{eqnarray}\label{eq:remote estimator}
    \hat{x}_{k} \eq \Eb[x_{k} | \IC_{k}^c] = A \Eb[x_{k-1} | \IC_{k}^c] + Bu_{k-1} 
\end{eqnarray}
with the initial value $
\hat{x}_0 = \bar{x}_0$, and $\IC_k^c$ denotes the remote information set. The control signal $u_k$ is generated according to control law $u_k =f(\IC_k^c)$, where $f \in \FC$ with $\FC$ being the admissible control law set. 
Then, the remote information set available for the controller is defined as $\IC_k^c = 
\{ z_{0:k}, \delta_{0:k-1},  a_{0:k}\} $ with the initial value $\IC_0^c = 
\{a_{0}\}$. 
Moreover, assume that the transmission success index $\delta_k$ will be returned to the local side with a one-step delay. 
Thus, the local information set available for the power scheduler by time $k$ is  $\IC_{k}^{p}= \{ x_{0:k},  \delta_{0:k-1},a_{0:k} \}$  with the initial value $\IC_{0}^{p} = \{  a_{0}\}$.

This article aims to co-design the power scheduler and control law to optimize control performance with limited transmission energy. More specifically, we consider both finite- and infinite-horizon problems.

\begin{problem}\label{Prob:finiteproblem}
(Finite-horizon problem) Find the optimal power scheduler $\pi^*$ and the control law $f^*$ by solving the following finite-horizon optimization problem:
\begin{eqnarray} \label{eq:finite opt}  
     \min_{ f,   \pi  }  & \Psi(f,\pi) =  J_c(f, \pi)  + \lambda J_p(\pi),
\end{eqnarray}
where the scalar $\lambda>0$ denotes the tradeoff multiplier. The control performance $J_c(f,\pi)$ is defined as 
\begin{eqnarray}\label{eq:finite LQG}
J_c(f,\pi) = \Eb\big[ \sum_{k=0}^{T-1 }  \hspace{-1em}& \gamma^k(x_{k}^\top Q x_{k} + u_{k}^\top R u_{k}) + \gamma^T x_{T}^\top Q_N x_{T} \big],  
\end{eqnarray} 
where $\gamma \in (0,1]$ is the discount factor.  The matrices $Q$, $Q_N$ are semi-definite positive and $R$ is definite positive, respectively. 
Assume that the pair $(A, Q^{\frac{1}{2}})$ is detectable, with $Q  = (Q^{\frac{1}{2}})^\top Q^{\frac{1}{2}}$.  Moreover, the total transmission energy consumption $J_p(\pi)$  is measured by $J_p(\pi) =   \sum_{k=0}^{T-1}\gamma^k p_{k} $. 
\end{problem}

\begin{problem}\label{Prob:infiniteproblem}
(Infinite-horizon problem) Find the optimal power scheduler $\pi^*$ and the control law $f^*$ by solving the following infinite-horizon optimization problem: 
\begin{eqnarray} \label{eq:infinite opt} 
     \min_{ f,   \pi  }  & \hat{\Psi}(f,\pi) =  \hat{J}_c(f, \pi)  + \lambda \hat{J}_p(\pi)
\end{eqnarray}
with $\hat{J}_c(f, \pi)$ being the infinite-horizon LQG function: 
\begin{eqnarray}\label{eq:infinite LQG}
\hat{J}_c(f,\pi) = \Eb\big[ \sum_{k=0}^{\infty }  \hspace{-1em}& \gamma^k(x_{k}^\top Q x_{k} + u_{k}^\top R u_{k}) \big],  
\end{eqnarray}
where $\gamma \in (0,1)$, 
and the transmission energy consumption is $\hat{J}_p(\pi) =   \sum_{k=0}^{\infty}\gamma^k p_{k}  $. 
\end{problem}

The discount factor $\gamma$ reflects how immediate and future costs are weighted.  A higher discount factor places more weight on future costs and vice versa.

\section{Finite-horizon case}\label{sec:finite}
In this section, we develop the solution to Problem~\ref{Prob:finiteproblem}. We will decompose the co-design optimization problem and design the optimal control law in the following.


\subsection{Optimal control}
In stochastic control systems, the control action generally has a dual effect. This means that it, on the one hand, stabilizes the system; on the other hand, reduces the system uncertainty by improving the system state estimate given the knowledge of past control actions, see \cite{bar1974dual}. 
It is shown in \cite{bar1974dual} that the dual effect does not exist when the conditional state estimate based on available information is independent of past control actions. 
When addressing the state-based scheduling of networked control systems, the dual effect can be removed by letting the scheduling policy be independent of past control signals \cite{ramesh2011on}. 
Note that the stochastic optimization problem considered involves two decision-makers: the local scheduler and the remote estimator.
We first provide the following notion to facilitate the search for the structural results of the optimization problem.  

\begin{definition}(Dominating policy)
Denote  $\UC$ as the set of all admissible policy pairs $(f,\pi)$. Consider the cost function $\Psi$ defined in the corresponding problem.  A set of policy pairs $\UC' \subset \UC$ is called a dominating class of policies, if for any feasible $(f,\pi) \in \UC$, there exists a feasible $(f',\pi') \in \UC'$, such that $\Psi(f',\pi') \le \Psi(f,\pi )$. 
\end{definition}

The following lemma identifies which class of policies is dominating for the problem \eqref{eq:finite opt} when the attack energy is known. Based on it, the original co-design optimization problem \eqref{eq:finite opt} is decomposed into two subproblems, i.e., optimal control and optimal scheduling design.

\begin{lemma}\label{lemma:CE}
Consider an admissible scheduling policy set $\Pi$, in which function only depends on random variables $\{x_0,w_{0:k-1}, 
a_{0:k}\}$. 
Then the set $\UC^{\text{CE}} =\{f^{\ast},\pi)| \pi \in \Pi \}$ is a dominating class of policies, where $f^{\ast}$ is the certainty equivalence controller:
\begin{eqnarray}\label{eq:finite controller}
u_{k}^{\ast}= f^{\ast}( \IC_k^c)= - L_k  \Eb[x_{k} |\IC_k^c]
\end{eqnarray}
with $L_k  = \gamma (R+\gamma B^{\top}P_{k+1}B)^{-1}B^{\top}P_{k+1} A $, and $P_{k+1} $ is solved from algebraic Riccati equation \cite{astrom2006introduction}:
\begin{eqnarray}\label{eq:Riccati matrix}
P_k  \eq   Q + \gamma A^{\top}P_{k+1} A \nonumber \\ 
&&\hspace{-0.5em} - \gamma^2 A^\top P_{k+1}B(R  + \gamma  B^{\top}P_{k+1} B)^{-1}B^{\top}P_{k+1}A 
\end{eqnarray}
with $P_N = Q_N$. 
\end{lemma}
\textbf{Proof}. Assume that there exists a triggering law $\tilde{\pi} = \{\tilde{\pi}_{1},\tilde{\pi}_{2},\dots\}$ being the function of random variables $\{x_{0},w_{0:k-1},a_{0:k}\}$. Note that the information pattern of the power scheduler and the controller are nested, i.e., $\IC_{k}^c  \subset \IC_{k}^{p}$.
In addition, we can see that $\{x_0, w_{0:k-1}\}$ can be fully recovered from $x_{0:k}$ and $u_{0:k-1}$, which is inferred from $\IC_k^p$ accessible to the local scheduler. Therefore, there exists a policy pair $(f,\tilde{\pi})$ producing identical decision variables as $(f,\pi)$ almost surely, i.e., $\tilde{\pi}_k(x_{0}, w_{0:k-1},a_{0:k}) = \pi(\IC_{k}^{p})$ 
holds almost surely. In other words, there exists a policy pair $(f,\tilde{\pi})$ that achieves the same cost as  $(f,\pi)$. 
Denote the remote estimation error as $   e_k = x_{k} - \Eb[x_{k}\vert \IC_k^c] 
$. 
According to \eqref{eq:remote estimator}, we have  $  \hat{x}_{k+1} = A \hat{x}_k + Bu_k   +\delta_k w_k + (1-\delta_k)A\Eb[e_k|\IC_k^c,\delta_k=0]$. 
Under the control law $f$ that is symmetric with respect to innovation $w_k$, we have 
$\Eb[e_k|\IC_k^c,\delta_k=0] = 0$, see \cite{soleymani2022value}. 
Then the remote estimation error $e_k$ evolves as 
\begin{eqnarray}\label{eq:estimation error}
e_{k+1} = \delta_k w_k + (1-\delta_k)(Ae_k+w_k)  
\end{eqnarray}  
with the initial value $e_0 = x_0 - \hat{x}_0$. Substituting the algebraic Riccati equation \eqref{eq:Riccati matrix} into the cost function \eqref{eq:finite opt}, we have 
\begin{eqnarray}\label{eq:finite opt1}
  \Psi(f,\pi)\eq \Eb\big[x_{0}^{\top}P_0x_{0} + \sum_{k=0}^{T-1} \big( \gamma^{k+1} w_{k}^{\top}P_k w_{k} \nonumber \\ 
   &&\hspace{-2em} + \gamma^k(u_{k}+L_k x_{k})^{\top}\Lambda_k(u_{k}+L_k x_{k})   + \lambda \gamma^k p_k\big) \big] 
\end{eqnarray} 
with $\Lambda_k =R  + \gamma B^{\top}P_{k+1} B$. Note that the first, the second, and the last terms of \eqref{eq:finite opt1} are independent of the control policy $f$. 
Substituting $x_k$ with $\Eb[x_{k}\vert \IC_k^c] + e_k$, we have 
\begin{eqnarray*}\label{eq:conditional iota}
&& \Eb[(u_{k}+L_k x_{k})^{\top}\Lambda_k(u_{k}+L_k x_{k})] \nonumber \\ 
\eq \Eb[(u_k+L_k\Eb[x_k|\IC_k^c])^\top \Lambda_k(u_k+L_k\Eb[x_k|\IC_k^c])   \nonumber \\   
&& + 2 (u_k+L_k\Eb[x_k|\IC_k^c])^\top \Lambda_k L_k e_k   +    (L_k e_k )^\top \Lambda_k L_ke_k ]  \nonumber \\ 
\eq \Eb[ (u_k+L_k\Eb[x_k|\IC_k^c])^\top \Lambda_k(u_k+L_k\Eb[x_k|\IC_k^c]) \nonumber \\ 
&& + (L_k e_k )^\top \Lambda_k L_k e_k ],
\end{eqnarray*}
where the second equality follows from the tower property of conditional expectation, i.e., $ \Eb[(u_k+L_k\Eb[x_k|\IC_k^c])^\top \Lambda_k L_ke_k  ]  =   \Eb[(u_k+L_k\Eb[x_k|\IC_k^c])^\top \Lambda_k L_k\Eb[e_k|\IC_k^c]]$ and $\Eb[e_k|\IC_k^c]=0$. 
Similar to the proof in \cite{molin2013on,molin2014suboptimal}, $\Eb[(L_k e_k )^\top \Lambda_k L_k e_k ]$ is independent of control law $f$. This also implies that the dual effect does not exist.
Then the optimal controller minimizing \eqref{eq:finite opt1} results in the certainty equivalence controller \eqref{eq:finite controller}.  
Moreover, we have $\Psi(f,\pi) = \Psi(f,\tilde{\pi}) \ge \min_{f \in \FC} \Psi(f,\tilde{\pi}) = \Psi(f^\ast,\tilde{\pi}) = \Psi(f^\ast,\pi')$, 
where $\pi'$ depends on $\IC_k^p$, and the first and the last equalities follow from $\tilde{\pi}_k(x_{0}, w_{0:k-1},a_{0:k}) = \pi(\IC_{k}^{p})$. Namely, for any feasible $(f,\pi) \in \UC$, the set  $(f^\ast, \pi') \in \UC$ is a dominating class of policy pairs such that $\Psi(f,\pi) \ge \Psi(f^\ast, \pi')$.  
 \hfill $\blacksquare$
 
Lemma~\ref{lemma:CE} implies that the original co-design problem can be decomposed into two subproblems, i.e., the optimal control law design and the optimal power scheduling design depending only on primitive random variables, without loss of optimality.
Given that the attack energy is available to both the power scheduler and the remote estimator, the information pattern between the remote controller and the local scheduler is shown to be nested.  Moreover, the one-step delay setting does not impose restrictions as the local side can infer the packet arrival status according to the control signal $u_{k-1}$ and control policy $f^\ast$. 
More specifically, in the one-step delay setting, decision sequences process as $\cdots \rightarrow \{u_{k-1},\delta_k\} \rightarrow \{u_{k},\delta_{k+1}\} \rightarrow \cdots$, which preserves the nested property. Similarly,
in the delay-free case, the nested property of the information pattern can be obtained by specifying the decision sequence ordering,  such as '$\cdots \delta_k \rightarrow u_k \rightarrow \delta_{k+1} \rightarrow u_{k+1},\cdots$', as in \cite{molin2013on}.

\subsection{Optimal power scheduling}

The following theorem will develop the optimal power scheduler in the finite horizon scenario. 
\begin{theorem}\label{theorem:finite}
Consider the optimization problem \eqref{eq:finite opt} for system \eqref{eq:plant}.
Fix the optimal control law as the certainty equivalence controller \eqref{eq:finite controller}. Let 
\begin{eqnarray}\label{eq:finite optimal schedule}
q_{k}^{\ast} \eq   \mathop{\argmin}\limits_{q_{k} \in \QC}
 \{ g(e_k,a_k,q_k) + q_k \iota_k    \},
\end{eqnarray}
where the stage cost is
\begin{eqnarray}\label{eq:stage cost}
g(e_{k},a_k, q_{k}) = \lambda p(q_k,a_k) + \gamma q_k e_k^\top A^\top \Sigma_{k+1} Ae_{k} 
\end{eqnarray} 
with  $\Sigma_k = L_k^\top(R+\gamma B^{\top}P_{k+1}B)L_k$ and 
$p(q_k,a_k) =  \Big(Q_f^{-1}\Big(\frac{q_k}{2}\Big)\Big)^2\frac{a_k+\sigma^2}{\alpha}. $
Moreover,  
$\iota_k = \Eb[V_{k+1}(\IC_{k+1}^p)|\IC_k^p,\delta_k = 0]-\Eb[V_{k+1}(\IC_{k+1}^p)|\IC_k^p,\delta_k = 1]$ and 
\begin{eqnarray}\label{eq:min Vk}
 V_{k}(\IC_k^p) =\min_{q_k \in \QC} \Eb[ \gamma^k g(e_k,a_k,q_k) +V_{k+1}(\IC_{k+1}^p) |\IC_k^p] ,  
\end{eqnarray}
for $k \in \mathbb{N}_{[0,T-1]}$ with the initial condition $V_{T}(\IC_{T}^p) = 0$. Then the optimal power scheduler $\pi^*(\IC_k^p)$ is given by  \begin{eqnarray}\label{eq:pk expression finite}
    p_k = \pi^*(\IC_k^p)=p(q_k^\ast,a_k).   
\end{eqnarray}
\end{theorem}  
\textbf{Proof}.
Substituting the optimal controller  \eqref{eq:finite controller}
into \eqref{eq:finite opt1} yields 
\begin{eqnarray}\label{eq:finite opt2}
   \Psi(f^\ast,\pi) \eq  \Eb\big[x_{0}^{\top}P_0x_{0} + \sum_{k=0}^{T-1} \big( \gamma^{k+1} w_{k}^{\top}P_k w_{k} \nonumber \\ 
   && \hspace{3em}+ \gamma^k e_k^{\top}\Sigma_k e_k   +\lambda \gamma^k p_k \big)\big] .
\end{eqnarray} 
Taking the conditional expectation of $e_{k+1}^{\top}\Sigma_{k+1} e_{k+1}$ with respect to  $\IC_k^p $, and by \eqref{eq:estimation error}, 
we have 
\begin{eqnarray}\label{eq:conditional e_k+1}
   && \Eb[   e_{k+1}^{\top}\Sigma_{k+1} e_{k+1} |\IC_k^p ] \nonumber \\  
  \eq \Eb\Big[          (1-\delta_k) (Ae_k+w_k)^\top \Sigma_{k+1} (Ae_k+w_k) \nonumber  \\ 
  && + \delta_k w_k^\top \Sigma_{k+1} w_k  \big|\IC_k^p \Big] \nonumber \\ 
\eq \Eb\Big[          (1-\delta_k) e_k^{\top}A^\top\Sigma_{k+1} A e_k  +  w_k^\top \Sigma_{k+1} w_k  \big|\IC_k^p \Big]   \nonumber \\ 
\eq \Eb\Big[    q_k e_k^{\top}A^\top\Sigma_{k+1} Ae_k   \big|\IC_k^p \Big] + \text{tr}(\Sigma_{k+1}W),
\end{eqnarray}
where the second equality establishes as $\Eb[w_k|\IC_k^p]=0$ and $w_k$ is independent of $e_k$,  the last equality follows from $\Eb[\Eb[(1-\delta_k)|\IC_k^p]\IC_k^p] = \Eb[q_k|\IC_k^p]$.
Then the original optimization problem \eqref{eq:finite opt} is reduced to 
\begin{eqnarray}\label{eq:simplified opt}
    \min_{\pi \in \Pi} \sum_{k=0}^{T}  \Eb\Big[ \gamma^k g(e_k,a_k, q_k) \big| \IC_k^p  \Big]. 
\end{eqnarray}
where $g(e_k,a_k, q_k)$ is defined in \eqref{eq:stage cost}. We omit the remaining terms of \eqref{eq:finite opt2} and the last term of \eqref{eq:conditional e_k+1} as they are independent of $\pi$. 
Note that $e_k$  can be fully recovered by the local power scheduler as $\delta_{k-1} \in \IC_k^p$.  
Let us apply the dynamic programming approach to \eqref{eq:simplified opt}. Synthesizing the optimal scheduler boils down to solving the value function $V_0(\IC_0^p)$, as defined in \eqref{eq:min Vk}. 
Moreover, 
\begin{eqnarray}\label{eq:V_k+1}
\hspace{-2em}&&\Eb[V_{k+1}(\IC_{k+1}^p)|\IC_k^p] 
= \Eb\Big[ q_k\Eb[V_{k+1}(\IC_{k+1}^p)|\IC_k^p,\delta_k = 0] \nonumber \\
\hspace{-2em}&&\hspace{6em} +  (1-q_k)\Eb[V_{k+1}(\IC_{k+1}^p)|\IC_k^p,\delta_k = 1]\Big],
\end{eqnarray}  
which follows from the law of total expectation. 
Substituting \eqref{eq:V_k+1} into \eqref{eq:min Vk} yields \eqref{eq:finite optimal schedule}, which gives the optimal power scheduler in \eqref{eq:pk expression finite}.  
\hfill $\blacksquare$

\begin{remark}\label{remark:greedy}
Theorem~\ref{theorem:finite} characterizes the optimal power scheduler for the finite-horizon problem \eqref{eq:finite opt}.  Note that it is difficult to compute $\iota_k$ in \eqref{eq:finite optimal schedule}, in particular for a long horizon $T$. An alternative greedy way to approximate $q^*_k$ is letting
\begin{eqnarray}\label{eq:greedy schedule}
  q_{k}^{\ast} \eq 
\mathop{\argmin}\limits_{q_{k} \in \QC}
 g(e_k,a_k,q_k).   
\end{eqnarray}
The greedy scheduling policy solved from \eqref{eq:greedy schedule} is suboptimal yet has a rather low computation complexity. 
\end{remark}

Lemma~\ref{lemma:CE} and Theorem~\ref{theorem:finite} assume that the attack energy $a_k$ is known, the following result considers the case when attack energy $a_k$  is unknown. 
\begin{lemma}
Consider the optimization problem \eqref{eq:finite opt} for system \eqref{eq:plant}. 
If the attack energy $a_k$ is unknown, consider an admissible scheduling policy set $\Pi'$, which only depends on random variables $\{x_0,w_{0:k-1}\}$.  
Then the set $\UC^{\text{CE}'} =\{f^{\ast},\pi)| \pi \in \Pi' \}$ with $f^\ast$ given by \eqref{eq:finite controller} is a dominating class of policies.
\end{lemma} 
\textbf{Proof}. 
Note that removing the attack energy $a_k$ from the information sets $\IC_k^p$ and $\IC_k^c$ does not alter the information structure between the local power scheduler and the remote controller.  
It follows the proof of Lemma~\ref{lemma:CE}.
Then the set $\UC^{\text{CE}'}$ constituted by a power scheduling $\pi \in \Pi'$ and the certainty equivalence controller \eqref{eq:finite controller} is a class of dominating policies.  \hfill $\blacksquare$

Next, we provide an approximation to the optimal scheduler when the attack energy $a_k$ is unknown. It follows from the proof of Theorem~\ref{theorem:finite}. Then, we consider the optimization problem \eqref{eq:simplified opt}. Taking the expectation of $p_k$ over the distribution $\DC_a$ yields  $ \Eb_{a_k \sim \DC_a} [p(q_k,a_k)] $. 
Then we replace $p_k$ with $ \Eb_{a_k \sim \DC_a} [p(q_k,a_k)]$ in scheduling policy \eqref{eq:finite optimal schedule} and \eqref{eq:stage cost}. Then, an approximation-based greedy scheduling is
 \begin{eqnarray}\label{eq:exp greedy schedule}
  q_{k}^{\ast} = \mathop{\argmin}\limits_{q_{k} \in \QC}
 \{ \mathop{\Eb}\limits_{a_k \sim \DC_a} [ \lambda p(q_k,a_k)]+ \gamma q_k e_k^\top A^\top \Sigma_{k+1} Ae_{k}  \}.  
\end{eqnarray}
\begin{remark}\label{remark:compare 2 greedy}
The greedy policy \eqref{eq:greedy schedule} requires real-time knowledge of the attack energy $a_k$ and does not rely on any distributional information about the attack energy, i.e., conditions 1) and 3) in Assumption~\ref{ass:attack property},  whereas \eqref{eq:exp greedy schedule} works the other way around. 
Furthermore, since the $p(q_k,a_k)$ given in \eqref{eq:stage cost} is linear in $a_k$, the cost function \eqref{eq:simplified opt} achieved by the greedy scheduler \eqref{eq:greedy schedule} remains consistent across various attack energy distributions with the same mean.  Additionally, since the average of greedy strategy \eqref{eq:greedy schedule} with respect to the random variable $a_k$ is identical to the approximation-based greedy strategy \eqref{eq:exp greedy schedule}, the performance of strategy \eqref{eq:exp greedy schedule} converges to that of strategy \eqref{eq:greedy schedule} as the time horizon $T$ increases. This finding will be demonstrated in the Simulation section. 
\end{remark}

\section{Infinite-horizon case}\label{sec:infinite}
In this section, we consider  the infinite-horizon problem, i.e., Problem~\ref{Prob:infiniteproblem}. 
Since the pair $(\sqrt{\gamma}A,B)$ is controllable, the algebraic Riccati equation
\begin{eqnarray}\label{eq:constant ARE}
    P =   Q + \gamma A^{\top}P A   - \gamma^2 A^\top PB(R  + \gamma  B^{\top}P B)^{-1}B^{\top}PA 
\end{eqnarray}
has a unique and positive semi-definite solution $P$. Accordingly, matrices $L_k$ $\Lambda_k$, $\Sigma_k$ are written as $L$, $\Lambda$ and $\Sigma$.  
The following assumption is essential for characterizing the optimal stationary scheduler.   
\begin{assumption}\label{assumption:stable}
    The minimum achievable packet dropout probability under the attack energy $a$, i.e., $q_{m}(a) =  2 Q_f\bigg( \sqrt{ \frac{\alpha p_{\max}}{a + \sigma^2}}\bigg)$, satisfies $    \Eb_{a\sim \DC_a}[q_{m}(a)] < \frac{1}{\rho(A)^2}.$ 
\end{assumption}


Similar to Lemma~\ref{lemma:CE}, we decompose the optimization problem \eqref{eq:infinite opt} into two subproblems as follows. 
\begin{lemma}
The set $\UC^{\text{CE}} =\{f^{\ast},\pi)| \pi \in \Pi \}$ is a dominating class of policies,  where $f^{\ast}$ is given by the certainty equivalence controller: 
\begin{eqnarray}\label{eq:inf controller}
u_{k}^{\ast}= f^{\ast}( \IC_k^c)= - L  \Eb[x_{k}\mid \IC_k^c]
\end{eqnarray}
with $L  =  \gamma (R+\gamma B^{\top}PB)^{-1}B^{\top}P A $. 
\end{lemma}
\textbf{Proof}. It follows from the proof of Lemma~\ref{lemma:CE}.  \hfill $\blacksquare$

Next we consider how to synthesize the optimal power scheduler. From \eqref{eq:finite opt2}, the cost function under the optimal control policy \eqref{eq:inf controller} is rewritten as 
\begin{eqnarray}\label{eq:cost optimal control}
    \hat{\Psi}(f^\ast,\pi) = \Eb\big[ \sum_{k=0}^{\infty}  \gamma^{k+1} w_{k}^{\top}P w_{k} + \gamma^k e_k^\top \Sigma e_k + \lambda \gamma^k p_k \big].
\end{eqnarray}
Accordingly, the optimization problem \eqref{eq:infinite opt} is reduced to
\begin{eqnarray}\label{eq:inf simplified opt}
    \min_{\pi \in \Pi} \sum_{k=0}^{\infty}  \Eb\Big[ \gamma^k g(e_{k},a_k, q_{k}) |\IC_k^p \Big] 
\end{eqnarray}
with the per-stage cost $g$ defined in \eqref{eq:stage cost}, $\Sigma = L^{\top}(R  + \gamma B^{\top}P B) L$ and initial value $(e_0,a_0):= (e,a)$.   
We next formulate the optimization problem \eqref{eq:inf simplified opt} as a MDP with the control model $(\XC,\QC,\PC,g)$, 
where $\XC:=\mathbb{R}^n \times \SC$ denotes the state space, 
$\QC$ denotes the action space, $\PC$ denotes the Borel measurable transition kernel defined on $(\XC,\QC)$.  
Note that $w_k$ obeys a distribution of $\NC(\mathbf{0},W)$. 
Denote $e^+,a^+$ as the next states of $e$ and $a$, respectively. Then, by \eqref{eq:estimation error}, the transition probability from $(e,a,q)$ to $(e^+,a^+)$ is given by 
\begin{eqnarray}\label{eq:transition}
&& \PC(e^+,a^+ \vert e,a, q ) \nonumber \\ 
\eq [ (1-q ) \mu_{w}(e^+)   + q \mu_{w}(e^+-Ae )] \DC_a(a^+),  \label{eq:decision based pdf}
\end{eqnarray} 
where 
$\mu_{w}$ is the p.d.f. of the random variable $w \sim  \NC(\mathbf{0},W)$. Note that \eqref{eq:transition} holds as the process noise $w$ is independent of attack energy $a \in \DC_a$ and $q$ is the current decision.

According to the optimization problem \eqref{eq:inf simplified opt}, define an associated $n$-stage cost under the policy $\pi \in \Pi$: $ J_n(e,a,\pi) := \Eb^\pi\big[ \sum_{t=0}^{n-1}\gamma^tg(e_t,a_t,q_t) \big] $
with the initial value $(e,a) \in \XC$ and $n \ge 1$. The decision variables $q_t$ for $t\in \mathbb{N}_{[0,n-1]}$ is chosen according to the policy $\pi$.  Let $G(e,a)$ be a class of nonnegative and lower semi-continuous (l.s.c.) functions on $(e,a) \in \XC$. Define the value iteration sequence $V_n(e,a) \in G(e,a)$, for $n \ge 1$ and $V_0(\cdot,\cdot) = 0$: 
\begin{eqnarray*}\label{eq:value iteration}
  V_n(e,a)\eq\min_{q \in \WC(e,a) }\big\{ g(e,a,q) + \gamma \Eb[ V_{n-1}(e^+,a^+|e,a,q)]\big\},
\end{eqnarray*}
where $\WC(e,a)$ highlights that, for any $ (e,a) \in \XC$, a non-empty set is associated to $(e,a)$. 
Then, we have 
$    V_{n }(e,a) = \inf_{\pi} J_n(e,a,\pi) $ 
given the initial value $(e,a) \in \XC$,
 for $n \ge 1$.

The following proposition shows that there exists an optimal stationary policy for the optimization problem \eqref{eq:inf simplified opt}. 
\begin{proposition}\label{proposition:MDP}
Let Assumptions~\ref{ass:attack property} and \ref{assumption:stable} hold, we have the following claims: 
\begin{enumerate}
    \item $\lim_{n\rightarrow\infty}V_n =V^\ast$ with $V^\ast = \inf_{\pi}\lim_{n\rightarrow\infty}J_{n}(e,a,\pi)$; 
    \item $V^\ast$ satisfies the Bellman optimality equation:  \begin{eqnarray}\label{eq:bellman}
      \hspace{-2em}  V^\ast(e,a) = \min_{q \in \WC(e,a)}\Big[ g(e,a,q)  + \gamma   \Eb[V^\ast(e^+,a^+)|e,a,q] \Big].
    \end{eqnarray} 
\item 
There exist a optimal stationary policy $\pi^\ast \in \Pi: \XC \rightarrow \QC$ minimizing the right-hand side of \eqref{eq:bellman} for all $(e,a) \in \XC$, i.e., $V^{\ast}(e,a)  =    g(e,a,\pi^\ast ) + \gamma \Eb[V^\ast(e^+,a^+)|e,a,\pi^\ast]$.  
 \end{enumerate}
\end{proposition}
\textbf{Proof}.  
According to \cite{hernandez1992discrete}, we need to verify the following conditions: 
\begin{enumerate}
    \item $g(e,a,q)$ is nonnegative, l.s.c. and inf-compact on $\XC \times \QC$; 
    \item the transition law $\PC$ is weakly continuous; 
    \item the multifunction $(e,a) \rightarrow \WC(e,a) $ is l.s.c.; 
    \item there exists a policy $\hat{\pi}$ such that $J_{\infty}(e,a,\hat{\pi}) :=\sum_{k=0}^{\infty}  \Eb^{\hat{\pi}}\Big[ \gamma^k g(e_{k},a_k, \hat{q}) \Big]  <\infty$ for each $(e,a) \in \XC$, where $\hat{q} = \hat{\pi}(\mathcal{I}_k^p)$. 
\end{enumerate}
The first condition holds by the definition of $g(e,a,q)$, as in \eqref{eq:stage cost}. It is inf-compact on $\XC \times \QC$ as the set 
$\{q \in \WC(e,a)| g(e,a,q) \le r \}$ is compact. The second condition holds as $\PC(e,a,q)$ is continuous on $(e,a,q) \in \XC \times \AC$. Condition 2) implies that, for any continuous and bounded function $V^\ast(e,a)$ on $(e,a,q)$, the map $(e,a,q) \rightarrow \int_{\XC}V^{\ast}(e,a) \PC(e^+,a^+|e,a,q){\rm d}e^+ {\rm d}a^+$ is continuous on $(e,a,q) \in \XC \times \AC$. 
The third condition holds as $\XC \times \QC$ is convex, see \cite{hernandez1994monotone}.  

To verify the last condition, we choose $\hat{\pi}(\IC_k^p) = p_{\max}$ for all $k$. Then, the expected dropout probability under $\hat{\pi} $ is $ \hat{q} = \Eb_{a\sim \DC_a}[q(a,p_{\max})]$ with $q(a,p_{\max}) =  2Q_f \Big(\sqrt{\frac{\alpha p_{\max}}{a+\sigma^2}} \Big)$.  
The next is to calculate $J_{\infty}(e,a,\hat{\pi})$. 
Define $  \theta_t^w :=  \tr \big(\sum_{r=0}^{t-1}(A^r)^\top \Sigma A^r
     W \big)$, $ \theta_t^x  := \tr \big( (A^t)^\top \Sigma A^tX_0\big) + \theta_t^w,  \quad {\rm for}~ t \ge 1.$
When $k=0$, 
$ \Eb^{\hat{\pi}}[\gamma^k  e_k^\top \Sigma e_k] = \tr(\Sigma X_0)$.
When $k = 1$, $\Eb^{\hat{\pi}}[\gamma^k  e_k^\top \Sigma e_k]   =  \gamma(\hat{q}\theta_1^x +(1-\hat{q})\theta_1^w) = \gamma\tr(\Sigma W + \hat{q}A^\top \Sigma AX_0 ).  $
When $k=2$, $\Eb^{\hat{\pi}}[\gamma^k  e_k^\top \Sigma e_k]   = \gamma^2 \big( (1-\hat{q})\theta_1^w+\hat{q}(1-\hat{q})\theta_2^w + \hat{q}^2 \theta_2^x \big) = \gamma^2 \tr(\Sigma W + \hat{q}A^\top \Sigma AW + \hat{q}^2(A^2)^\top\Sigma A^2 X_0). $
By induction, for all $k\ge 1$, we have 
\begin{eqnarray}\label{eq:stage repara}
\hspace{-1em}&& \Eb^{\hat{\pi}}[\gamma^k  e_k^\top \Sigma e_k] \nonumber \\ 
\eq \gamma^k\tr\Big(\hat{q}^k (A^k)^\top\Sigma A^k X_0 + \sum_{r=0}^{k-1}\hat{q}^r(A^r)^\top\Sigma A^r W   \Big).  
\end{eqnarray}
Sum \eqref{eq:stage repara} over all $k \ge 0$, we obtain 
\begin{eqnarray}\label{eq:cost bound 1}
\hspace{-2em}&& \Eb^{\hat{\pi}}\bigg[\sum_{k=0}^{\infty} \gamma^k  e_k^\top \Sigma e_k\bigg]    
= \frac{\gamma}{1-\gamma}\tr \bigg(\sum_{r=0}^{\infty} (\gamma\hat{q})^r(A^r)^\top\Sigma A^r W \bigg)\nonumber \\ 
\hspace{-2em}&& \hspace{9em} +   \tr \bigg( \sum_{r=0}^{\infty}(\gamma\hat{q})^r (A^r)^\top \Sigma A^r X_0  \bigg). 
\end{eqnarray}
When $\gamma\hat{q}(\rho(A))^2 < 1$, we have $\sum_{k=0}^{\infty}\Eb^{\hat{\pi}}[\gamma^k  e_k^\top \Sigma e_k] < \infty$. Then Assumption~\ref{assumption:stable} is sufficient for its establishment. Moreover, we have 
\begin{eqnarray}\label{eq:cost difference}
&& \Eb^{\hat{\pi}}\bigg[\sum_{k=0}^{\infty}\gamma^k  e_k^\top \Sigma e_k\bigg] -\Eb^{\hat{\pi}}\bigg[\sum_{k=0}^{\infty} \gamma^{k+1}  \hat{q}_k e_k^\top A^\top \Sigma Ae_k\bigg]        \nonumber \\ 
 \eq \tr(\Sigma X_0) + \frac{\gamma}{1-\gamma} \tr(\Sigma W ), 
\end{eqnarray}
which holds by substituting \eqref{eq:estimation error} into \eqref{eq:cost difference}. 
The total transmission cost under the power scheduling law $\hat{\pi}$ is 
\begin{eqnarray}\label{eq:power bound}
J_p^{\hat{\pi}} =    \sum_{k=0}^\infty \gamma^k p_{\max} = \frac{p_{\max}}{1-\gamma}. 
\end{eqnarray}
Combining  \eqref{eq:cost bound 1}, \eqref{eq:cost difference} and \eqref{eq:power bound}, we obtain that   
$$J_{\infty}(e,a,\hat{\pi}) = \Eb^{\hat{\pi}}\bigg[\sum_{k=0}^{\infty} \gamma^{k+1} \hat{q}_k e_k^\top A^\top \Sigma Ae_k\bigg] + \lambda J_p^{\hat{\pi}} < \infty.$$ 
According to \cite{hernandez1992discrete}, claims 1-3) holds.  \hfill $\blacksquare$

The following theorem develops the optimal scheduler and analyzes its performance. 
\begin{theorem}
Consider the optimization problem \eqref{eq:infinite opt} for system \eqref{eq:plant}. 
Fix the optimal control law as the certainty equivalence controller \eqref{eq:inf controller}.  Let Assumptions \ref{ass:attack property} and \ref{assumption:stable} hold and let $$q^\ast (e,a) = \mathop{\argmin}_{q \in \QC} \big\{ g(e,a,q)  + \gamma \Eb[V^\ast(e^+,a^+)|e,a,q]   \big\},$$  
where $V^\ast(e,a)$ is solved from Bellman equation \eqref{eq:bellman}.
\begin{enumerate}
    \item  The optimal power scheduler $ \pi^*$ is given by 
\begin{eqnarray}\label{eq:pk expression infinite}
p=  \pi^\ast(e,a)  = \Big(Q_f^{-1}\Big(\frac{q^\ast(e,a)}{2}\Big)\Big)^2\Big(\frac{a+\sigma^2}{\alpha}\Big). 
\end{eqnarray}
\item 
Let $\tilde{q}$ be the solution of 
\begin{eqnarray}\label{eq:optimal const scheduler} 
 \hspace{-2em}&& \argmin_{q \in \QC}  \bigg\{ \frac{\gamma\tr(  \Theta W) }{1-\gamma} + \tr( \Theta X_0)+ \frac{ \lambda \Eb_{a \sim \DC_a} [p(q,a)]}{1-\gamma} \bigg\}, \nonumber  \\ 
\hspace{-2em}&& {\mathrm{s.t.}}~\gamma q A^\top \Theta A  + \Sigma = \Theta.
\end{eqnarray}
The total cost achieved by the optimal power scheduler and the optimal control is upper-bounded by 
\begin{equation}\label{eq:cost upperbound}
 \hat{\Psi}(f^\ast,\pi^\ast)  \le \frac{\gamma\tr( PW+\tilde{\Theta} W) +\lambda \tilde{p}}{1-\gamma}   + \tr(\tilde{\Theta} X_0),   
\end{equation}
where $\tilde{p} = \mathop{\Eb}\limits_{a_k \sim \DC_a} [ p(\tilde{q},a_k)]$ and
$\tilde{\Theta} $ satisfies 
\begin{equation}\label{eq:lyapunov equation}
\gamma \tilde{q} A^\top \tilde{\Theta} A  + \Sigma = \tilde{\Theta}.
\end{equation}
\end{enumerate}
\end{theorem}
\textbf{Proof.} 
The first claim follows from Proposition~\ref{proposition:MDP}. Next, we show the second claim.  Define a class of constant-power scheduling law $\bar{\Pi}$ and denote $ \tilde{\pi} \in \bar{\Pi}$ as the optimal constant-power scheduling law minimizing the total cost \eqref{eq:cost optimal control}. Given Assumption~\ref{assumption:stable}, 
\begin{eqnarray}\label{eq:lyapunov sum}
\tilde{\Theta} = \sum_{k=0}^{\infty}(\gamma\tilde{q})^k 
(A^k)^\top \Sigma A^k.  
\end{eqnarray} 
is the solution to Lyapunov equation \eqref{eq:lyapunov equation}. 
Substituting \eqref{eq:lyapunov sum}  into \eqref{eq:cost bound 1}, we have
\begin{eqnarray}\label{eq:cost bound 2}
  \Eb^{\tilde{\pi}}\bigg[\sum_{k=0}^{\infty} \gamma^k  e_k^\top \Sigma e_k \bigg] \eq \frac{\gamma}{1-\gamma} \tr(\tilde{\Theta} W ) + \tr(\tilde{\Theta}X_0).
\end{eqnarray} 
Substituting \eqref{eq:cost bound 2}  and $J_p^{\tilde{\pi}} =    \sum_{k=0}^\infty \gamma^k \tilde{p} = \frac{\tilde{p}}{1-\gamma} $  into \eqref{eq:cost optimal control}, we have  
\begin{align*}
 & \hspace{1em} \hat{\Psi}(f^\ast,\tilde{\pi}) \nonumber \\ 
  &= \Eb^{\tilde{\pi}}\bigg[\sum_{k = 0}^{\infty} \gamma^{k+1}w_k^\top P w_k \bigg]    + \frac{\gamma \tr(\tilde{\Theta} W)}{1-\gamma} +\tr(\tilde{\Theta} X_0)  + \frac{\lambda \tilde{p}}{1-\gamma} \nonumber \\  
 &= \frac{\gamma\tr( PW+\tilde{\Theta} W) + \lambda \tilde{p}}{1-\gamma}   + \tr(\tilde{\Theta} X_0). 
\end{align*}
Thus,  $\tilde{p} = \tilde{\pi}(e,a)$ solved from \eqref{eq:optimal const scheduler} is the optimal constant power, where $\frac{\gamma{\mathrm{tr}}(PW)}{1-\gamma}$ is omitted as it is independent of scheduling policy. Note that the scheduling policy \eqref{eq:pk expression infinite} is optimal among the class of policies depending on primitive random variables described by $\IC_k^p$, including $\bar{\Pi}$. As a consequence, the performance achieved by the scheduling policy $\pi^\ast$ is upperbounded by  that achieved by $\tilde{\pi}$, i.e., $\hat{\Psi}(f^\ast,\pi^\ast) \le \hat{\Psi}(f^\ast,\tilde{\pi})$. 
Thus, we obtain the second conclusion. 
   \hfill $\blacksquare$
   
Since Problem \ref{Prob:infiniteproblem} considers discounted criteria, the minimum expected cost \eqref{eq:cost upperbound} is affected by the attack energy distribution $\DC_a$, the discount factor $\gamma$, the covariance of state initial value $X_0$. This aligns with the result of the discounted optimization problem, see \cite{hernandez1992discrete}.

\section{Simulation}\label{sec:simulation} In this section,  we provide numerical examples to illustrate our results. 
Consider a second-order system with system dynamics $A = {\rm diag}\{1.3,-1.1\}$, $B = [0.1~0.1]^\top$, process noise covariance $W=\mathrm{diag}\{0.001,0.001\}$. The weighting matrices in the LQG function \eqref{eq:finite LQG} are chosen as $Q = R = {\rm diag}\{1,1\}$. The discount factor is chosen as $\gamma = 0.9$. The covariance of the initial value is chosen as $X_0 = {\rm diag}\{0.01,0.01\}$. The communication channel parameters are chosen as $\sigma^2=1$, $\alpha = 3$, see \cite{li2016sinr}. The attack energy is chosen as a uniformly distributed random variable between $[0,1]$, i.e., $a_k \sim U[0,1]$.
We choose the time horizon as $T=100$ and the tradeoff multiplier as $\lambda = 1$. 
 \begin{figure}[t]
    \centering
    \includegraphics[width=0.4\textwidth]{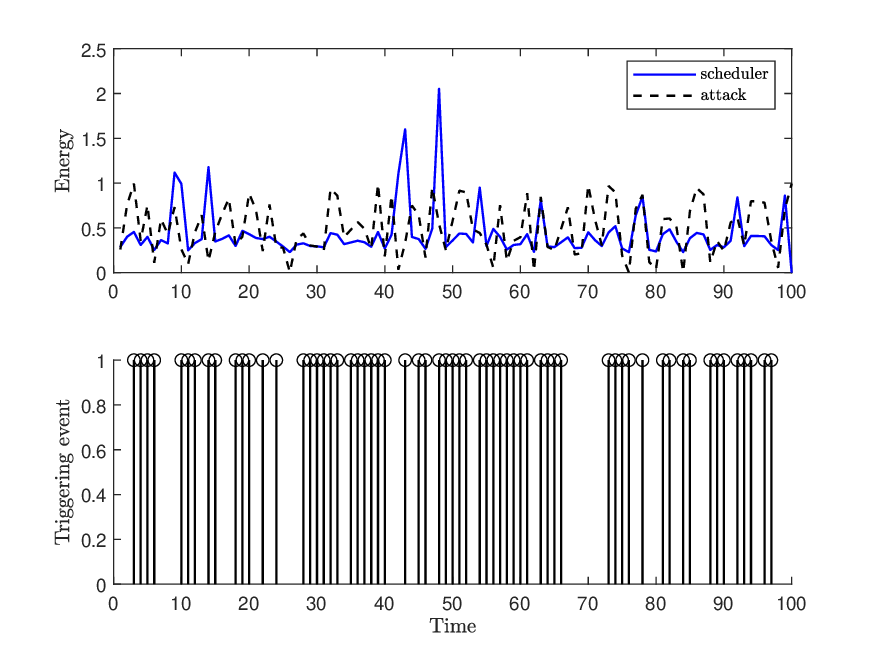}
    \centering 
    \caption{From top to bottom: attack energy and transmission power; transmission success index under greedy scheduler \eqref{eq:greedy schedule}.}
    \label{fig:scheduler}
\end{figure} 
\begin{figure}[t]
    \centering
    \includegraphics[width=0.4\textwidth]{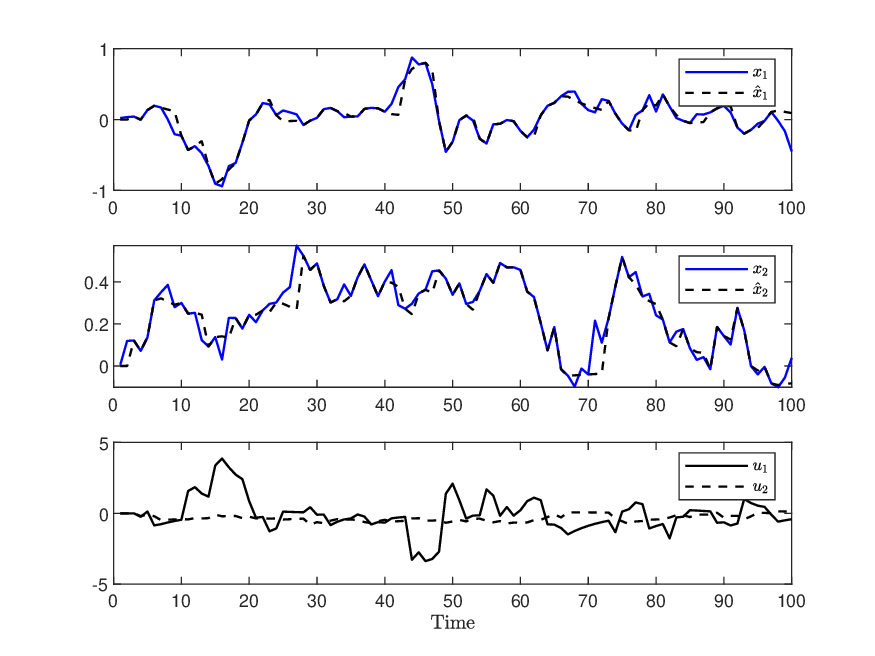}
    \centering 
    \caption{From top to bottom: system state; control signal under the greedy scheduler \eqref{eq:greedy schedule}.}
    \label{fig:dynamics}
\end{figure} 
\begin{figure}[t]
    \centering
    \includegraphics[width=0.4\textwidth]{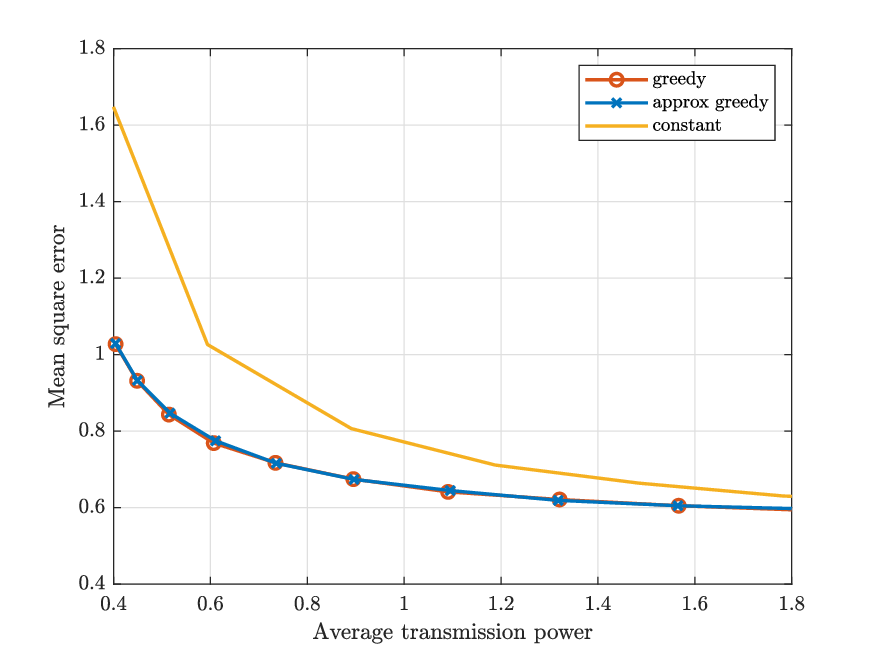}
    \centering 
    \caption{Tradeoff between mean square error and average transmission power archived by the greedy scheduler \eqref{eq:greedy schedule}, the approximation-based greedy scheduler \eqref{eq:exp greedy schedule}, and constant-power schedulers. }
    \label{fig:tradeoff}
\end{figure} 
\begin{figure}[t]
    \centering
    \includegraphics[width=0.4\textwidth]{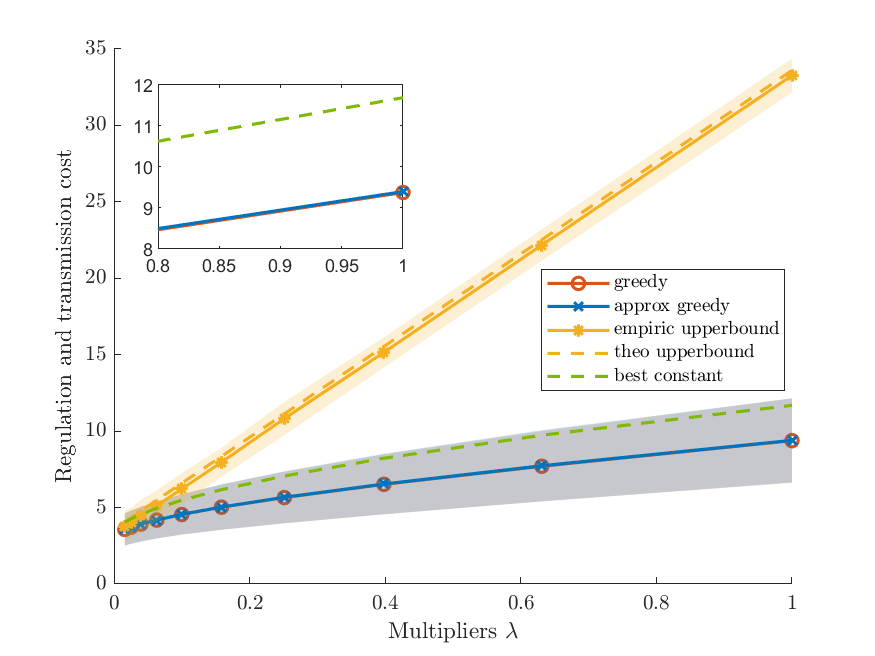}
    \centering 
    \caption{The total cost achieved by the greedy scheduler \eqref{eq:greedy schedule}, the approximation-based greedy scheduler  \eqref{eq:exp greedy schedule}. The theoretical and empirical upper bounds \eqref{eq:cost upperbound} of the total cost. The theoretical and empirical cost achieved by constant-power scheduler with $p_k=3$. }
    \label{fig:total_cost}
\end{figure} 

\begin{figure}[t]
    \centering
    \includegraphics[width=0.4\textwidth]{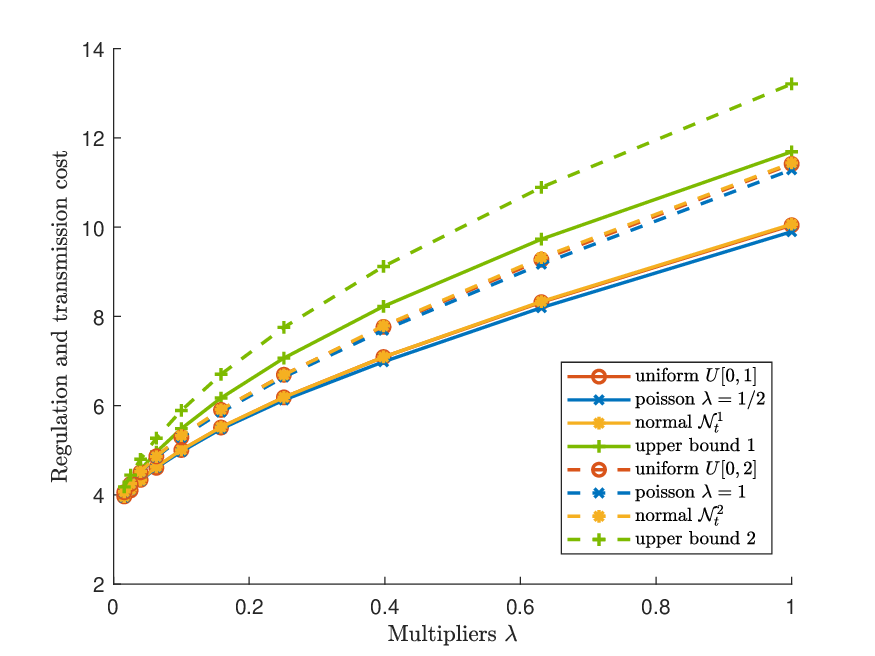}
    \centering 
    \caption{The empirical average cost achieved by the greedy scheduler \eqref{eq:greedy schedule} and the corresponding theoretical upper bounds \eqref{eq:cost upperbound} with different attack energy distributions. }
    \label{fig:compare noise}
\end{figure} 

The top subfigure of Fig.~\ref{fig:scheduler} depicts the attack energy and the transmission power determined by scheduler \eqref{eq:greedy schedule} for $k \in \mathbb{N}_{[0,T]}$. The bottom subfigure of Fig.~\ref{fig:scheduler} depicts the transmission success index for  $k \in \mathbb{N}_{[0,T]}$.  Fig.~\ref{fig:scheduler} shows that transmission success is affected by both the transmission power and attack energy. Additionally, a higher attack energy generally leads to higher transmission power.  Fig.~\ref{fig:dynamics} depicts the system state trajectory, its estimate, and control signal.  Fig.~\ref{fig:dynamics} shows that the remote estimator effectively tracks the system dynamics. 
Fig.~\ref{fig:tradeoff} depicts the mean square error achieved by the greedy schedulers  \eqref{eq:greedy schedule} and \eqref{eq:exp greedy schedule} and the constant-power schedulers, with the average transmission power $p = [0.4,1.8]$. The mean square error is measured by  $\frac{1}{T}\sum_{k=0}^T   e_k^\top A^\top \Sigma A e_k $ and the average transmission power is measured by $ \frac{1}{T}\sum_{k=0}^T   p_k$.  Fig.~\ref{fig:tradeoff} shows that for the same average transmission power, the performance of greedy schedulers  \eqref{eq:greedy schedule} approaches that of approximation-based scheduler \eqref{eq:exp greedy schedule}, which aligns with Remark~\ref{remark:compare 2 greedy}. Furthermore, they outperform the constant-power schedulers in achieving smaller mean square error. 

Fig.~\ref{fig:total_cost} depicts the total regulation and transmission costs \eqref{eq:finite opt2} with different multipliers under the greedy scheduler \eqref{eq:greedy schedule}, the approximation-based greedy scheduler \eqref{eq:exp greedy schedule}, and constant-power schedulers. We choose tradeoff multiples $\lambda \in [0.01,1]$. 
Monte Carlo simulation runs $20000$ trials. 
We choose $\sum_{k=0}^T \big(\gamma^k e_k^\top  \Sigma e_k +\lambda \gamma^k p_k\big)$ as the empirical total regulation and transmission cost.
For the empirical results, shaded areas represent $\pm$ one standard deviation over 20000 trials. 
A small shaded area means that the designed scheduling policy delivers steady performance. The theoretical upper bound of the cost, i.e., the cost achieved by the optimal constant-power scheduling $\tilde{\pi}$, is $\frac{\gamma\tr( \tilde{\Theta} W) + \lambda \tilde{p}}{1-\gamma}   + \tr(\tilde{\Theta} X_0)$, where $\tilde{p}$ and $\tilde{\Theta}$ are defined in \eqref{eq:lyapunov equation}.
The theoretical cost achieved by a transmission scheduler using constant power $p_k = p_{\max} =3$ is $\frac{\gamma\tr( \hat{\Theta} W) + \lambda p_{\max}}{1-\gamma}   + \tr(\hat{\Theta} X_0)$, where $\hat{\Theta}$ satisfies $\gamma \hat{q} A^\top \hat{\Theta} A  + \Sigma = \hat{\Theta}$.  
Fig.~\ref{fig:total_cost} shows that the theoretical costs $\hat{\Psi}(f^\ast,\tilde{\pi})$ and $\hat{\Psi}(f^\ast,\hat{\pi})$ both match their corresponding empirical costs, which illustrates the effectiveness of theoretical bound calculation.
Additionally, it shows that the greedy schedulers \eqref{eq:greedy schedule} and \eqref{eq:exp greedy schedule} both outperform arbitrary constant-power schedulers, including the optimal one. Furthermore, the greedy scheduler \eqref{eq:greedy schedule}  achieves the performance comparable to that of the approximation-based greedy scheduler \eqref{eq:exp greedy schedule}, echoing the statement in Remark~\ref{remark:compare 2 greedy}.

Fig.~\ref{fig:compare noise} depicts the total regulation and transmission costs achieved by greedy scheduler \eqref{eq:greedy schedule} and its theoretical upper bounds \eqref{eq:cost upperbound} under different attack distributions. We choose two uniform distributions $U_1[0,1]$ and  $U_2[0,2]$; two Poisson distributions $\mathcal{D}_{p}^{\lambda}$ with the p.d.f. of  $f(k,\lambda) = \mathrm{Pr}(X=k) = \frac{\lambda^k e^{-\lambda}}{k!}$, where $\lambda \in \{1/2,1\}$. Define the truncated normal distribution with p.d.f $ f(x,\mu,\sigma,a,b) = \frac{1}{\sigma}\frac{\varphi(\frac{x-\mu}{\sigma})}{\Phi(\frac{b-\mu}{\sigma}) - \Phi(\frac{a-\mu}{\sigma})}$ for $x\in[a,b]$ and $f = 0$ otherwise, where $\varphi$ is the p.d.f of the standard normal distribution and $\Phi(x) = \frac{1}{2}(1+{\mathrm{erf}}(x/\sqrt{2}))$.  We choose two truncated normal distribution $\mathcal{N}_t^1 $ and $\mathcal{N}_t^2 $ with $f(x,1/2,1/12,0,1)$ and $f(x,1,1/3,0,2)$. Note that the distributions $U_1$, $\mathcal{D}_p^{1/2}$ and $\mathcal{N}_t^1$ have the same mean value, and distributions $U_2$, $\mathcal{D}_p^{1}$ and $\mathcal{N}_t^2$ have the same mean value. Monte Carlo runs 20000 trials. For clarity, we only plot the mean value of the empirical costs and omit their standard deviations. 
Fig.~\ref{fig:compare noise} shows that, given the same average attack energy, the greedy scheduler achieves the same performance across various attack energy distributions. This observation aligns with Remark~\ref{remark:compare 2 greedy}. Moreover, under the same scheduling policy, the total cost increases as the attack energy increases.  
Additionally, theoretical upper bounds successfully constrain the empirical costs  under different attack energy distributions.

\section{Conclusion}\label{sec:conclusion}
In this article, we have studied the optimal co-design of control law and transmission power scheduler that minimizes the regulation and transmission costs for networked control systems under DoS attacks. Given the acknowledgment signal from the remote controller and same knowledge about the attack energy, the information structure between the controller and the power scheduler is nested. Then, we showed that the original co-design can be decomposed into the optimal control design, yielding a certainty equivalence controller, and the optimal power scheduling design, which is tracked by dynamic programming approaches. Expressions of the optimal power scheduling were provided in both finite- and infinite-horizon cases. To ease the computational complexity in finite-horizon dynamic programming, an alternative greedy scheduler was developed for implementation.  Additionally, in the infinite-horizon case, we provided the upper bound of the total regulation and transmission cost under the proposed scheduler.  Nevertheless, the proposed design relies on specific assumptions about the nature of DoS attacks, such as magnitude or distribution, which may limit its applicability in more diverse or unpredictable attack scenarios. As a result, its effectiveness may be limited when facing more sophisticated or unpredictable attack patterns. To address this limitation, future research will focus on developing adaptive power scheduling mechanisms that can dynamically adjust in response to real-time detection of evolving or unknown DoS attacks.  

\appendices



\

\bibliographystyle{ieeetr}
\bibliography{references}

\end{document}